\documentstyle[preprint,pra,aps]{revtex}

\begin{document}

\title{\bf Calculation of the positron bound state with the copper atom}

\author{V. A. Dzuba, V. V. Flambaum, G. F. Gribakin, and C. Harabati}

\address{School of Physics, The University of New South Wales, 
Sydney 2052, Australia}

\date{\today}
\maketitle

\tightenlines

\begin{abstract}
A new relativistic method for calculation of positron binding to atoms
is presented. The method combines a configuration interaction treatment
of the valence electron and the positron with a many-body perturbation theory
description of their interaction with the atomic core. We apply this method
to positron binding by the copper atom and obtain the binding energy of
170 meV ($\pm 10\%$). To check the accuracy of the method we use a similar
approach to calculate the negative copper ion. The calculated electron affinity
is 1.218 eV, in good agreement with the experimental value of 1.236 eV.
The problem of convergence of positron-atom bound state calculations
is investigated, and means to improve it are discussed. The relativistic
character of the method and its satisfactory convergence make it a suitable
tool for heavier atoms.
\end{abstract} 
\vspace{1cm}
%***************************************************************************
\narrowtext
\section{Introduction}

Bound states of positrons with neutral atoms have not been detected
experimentally yet. For a long time the prevailing view was that neutral
atoms do not bind positrons. For example, Aronson {\it et al}
\cite{aronson} proved that positron binding to hydrogen is not possible,
and Gertler {\em et al} \cite{gertler} showed that a ground-state
helium atom could not bind a positron. In a number of calculations
positron binding was observed for alkalis and second column atoms
\cite{clary,ward,szmyt}. However, important physical effects, such as virtual
or real positronium (Ps) formation, were neglected in those works. As a result,
the binding was largely considered as an artifact of the approximations used,
or the positron bound states found were unstable against Ps emission.
This situation has clearly changed now. Firstly, a many-body theory
calculation by Dzuba {\em et al.} \cite{Dzuba95} indicated that atoms with
larger dipole polarizabilities and ionization potentials greater then 6.8 eV
(Ps binding energy) can bind positrons, and predicted positron
binding energies for Mg, Zn, Cd and Hg. Subsequently, a number of recent
calculations have shown and even proved, for a few lighter atoms, that
positron-atom bound states do exist, \cite{Mitroy97,Strasburger98,small,%
Yuan98,Mitroy98,RyMitVar98,Mitroy98b,Mitroy98c,Mitroy99a,Mitroy99b}.

For the problem of positron-atom binding the atoms should be divided into two
groups: those with the ionization potential $I$ smaller than 6.8 eV, and those
with $I>6.8$ eV. For the former the lowest fragmentation threshold of the
positron-atom system is that of a positive ion and a Ps atom.
Consequently, positron binding to such atoms should rather be described
as binding of the Ps to the corresponding positive ion. Indeed, the
`ion $+$ Ps' component in their wave function is large, as shown by the
calculations for Li-$e^+$, Na-$e^+$ and He$\,2^3S$-$e^+$
\cite{Mitroy97,small,Yuan98,Mitroy98,RyMitVar98}. For atoms with $I>6.8$ eV
the positron-atom bound state is indeed an `atom $+$ $e^+$' system, at large
positron-atom separations. However, the process of virtual Ps formation in
this system is very important \cite{Dzuba95}, especially when $I$ is close
to 6.8 eV. This effect makes positron-atom bound states a strongly correlated
atomic system. The correlations in it are stronger than those one finds in its
electron analogues, atomic negative ions. This feature makes the
positron-atom bound complexes very interesting for the atomic theory. This also
makes them a challenging testing ground for applications of modern numerical
methods of atomic structure calculations.

% The most interesting candidates for positron-atom bound states are atoms
% which have big polarizabilities and big ionization potential. Ionization
% potential should be bigger than the energy of positronium (6.8 eV).
% This makes a positron-atom bound state stable against positronium emission.
% Big polarizability is needed because positron-atom interaction is roughly
% proportional to the polarizability of the atom \cite{Dzuba95}.
% Very few atoms which satisfy both these conditions were
% investigated so far. In our previous paper we demonstrated that 
% positron-atom bound states can exist for Mg, Zn, Cd and Hg \cite{Dzuba95}.
% Mitroy {\it et al} calculated energies of positron bound
% states with Be, Mg, Cu and Ag atoms \cite{Mitroy98,Mitroy98b,Mitroy98c}. 
% (There are also
% many calculations of positron interaction with atoms which have 
% small ionization potential \cite{Mitroy98,small}.)

The main difficulty in calculations of positron interaction with atoms
comes from the strong electron-positron Coulomb attraction which leads to
virtual positronium formation \cite{Dzuba95}. One can say that it gives rise
to a specific short-range attraction between the positron and the atom, in
addition to the usual polarizational potential which acts between a neutral
target and a charged projectile \cite{Dzuba:93,Gribakin:94,Dzuba:96}. This
attraction can not be treated accurately by perturbations and some all-order
technique is needed. In our earlier works \cite{Dzuba95,Gribakin:94,Dzuba:96}
we used the Ps wave function explicitly to approximate the virtual
Ps-formation contribution to the positron-atom interaction. The same physics
may also explain the success of the stochastic variation method in
positron-atom bound state calculations (see \cite{RyMitVar98} and Refs.
therein). In this approach the wave function is expanded in terms of
explicitly correlated Gaussian functions which include factors
$\exp (-\alpha r_{ij}^2)$ with interparticle distances $r_{ij}$.
Using this method
Ryzhikh and Mitroy obtained positron bound states for a whole range of atoms
with both $I<6.8$ eV (Li, Na, and He$\,2^3S$), and $I>6.8$ eV
(Be, Mg, Zn, Cu, and Ag). This method is well suited for few-particle
systems. Its application to heavier systems is done by considering the
Hamiltonian of the valence electrons and the positron in the model potential
of the ionic core. However, for heavier atoms, e.g., Zn, the calculation
becomes extremely time consuming \cite{Mitroy99a}, and its convergence cannot
be ensured.

Another non-perturbative technique is configuration interaction (CI) method
widely used in standard atomic calculations. This method has been applied
to the positron-copper bound state in \cite{Mitroy99b}. In this work the
single-particle orbitals of the valence electron and positron are chosen as
Slater-type orbitals, and their interaction with the Cu$^+$ core is
approximated by the sum of the Hartree-Fock and model polarization potentials.
The calculation shows slow convergence with respect to the number of spherical
harmonics included in the CI expansion, $L_{\max }=10$ being still not
sufficient to extrapolate the results reliably to $L_{\max }\rightarrow
\infty$.

% There is one more all-order technique which seems to be very
% suitable for the problem. This is so called coupled cluster (CC) 
% approach \cite{Lindgren}. The CC method includes double correlations 
% between every pair of atomic particles in all orders. This is exactly 
% what is needed for accurate description of positron-atom interaction. 
% CC method is widely used for accurate atomic calculations.
% However, we are unaware of any applications of the CC method 
% to the positron-atom problem.

In the present work we calculate the ground states of Cu-$e^+$ and Cu$^-$
systems using a CI calculation within a spherical cavity of finite radius $R$.
This procedure
facilitates the convergence of the CI expansion in the difficult positron-atom
case, and we show how to extrapolate the results to the $R\rightarrow \infty $
limit. The CI method which we use is based on the combined relativistic
configuration interaction and many-body perturbation theory method (CI+MBPT)
developed in our earlier work \cite{Kozlov} for precise calculations of
many-electron atoms with more than one valence electron. It was shown there
that correlations between the core and valence electrons are very important
and often contribute more to the energy than the correlations between the
valence electrons. The core-valence correlations are included into the
effective CI Hamiltonian of valence electrons by means of many-body
perturbation theory. This allows us to achieve high accuracy in calculations
of atomic energies and transition amplitudes. In the present work we adapt
this approach to the positron problem.

As a single-particle basis for the CI calculations we use $B$-splined
\cite{deBoor} Hartree-Fock wave functions in the cavity of finite radius
$R$. The $B$-spline technique has been successfully used in atomic
calculations 
for many years (see, e.g., review \cite{Sapirstein}) and has been recently
incorporated with the CI+MBPT method \cite{Johnson}. The use of $B$-splines
ensures good convergence of the CI calculation with respect to the number
of radial orbitals. Convergence is further controlled by varying the cavity
radius, while the effect of a finite cavity size on the energy of the system
is taken into account analytically.

We have chosen the copper atom for the positron bound-state calculations
for several reasons. First, this atoms looks like a good candidate for
positron-atom bounding. It has a large  polarizability of 40 a.u. \cite{pol},
and its ionization potential $I=7.724$ eV \cite{Moore} is not too far
from the Ps binding energy of 6.8 eV, which ensures a sizable contribution
of virtual Ps to the positron-atom attraction. Second, copper has a relatively
simple electronic structure with only one valence electron above closed shells.
This makes the positron-copper problem effectively a two-particle problem
well suited for application of the CI+MBPT method. Third, there are accurate
experimental data and a number calculations for the energy of the copper
negative ion. Thus, we can test our method on Cu$^-$ and compare the
results with those obtained by other techniques. Last but not least, the
existence of the positron-copper bound state was predicted by Ryzhik
and Mitroy \cite{Mitroy98b} in the framework of the stochastic variational
method, which allows us to compare the results obtained with the two
different techniques.

%------------------------------------------------------------------
\section {Method of calculation} \label{method}

\subsection{Effective Hamiltonian}

We use the relativistic Hartree-Fock method in the $V^{N-1}$ approximation
to obtain the single-particle basis sets of electron and positron orbitals
and to construct an effective Hamiltonian. The main point for this choice
is the simplicity of the MBPT, as discussed in Ref. \cite{Kozlov}.
The self-consistent potential is determined for the Cu$^+$ ion and
the single-particle states of the external valence electron and the positron
are  calculated in the field of the frozen core.

\begin{figure}[b]
\psfig{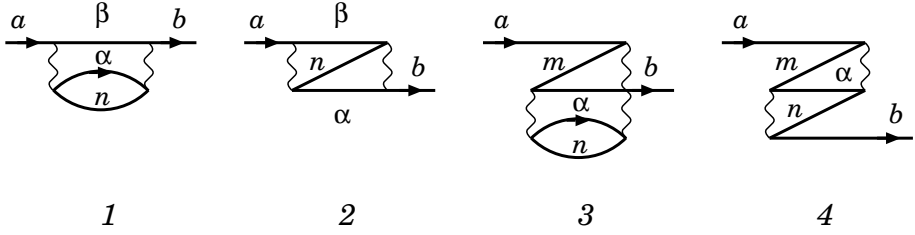}
%\epsfxsize=10cm
%\epsfbox{sigmae.eps}
\caption{Second-order diagrams for the self-energy of the valence 
electron ($\hat \Sigma_e$ operator). Summation of excited electron states
$\alpha $ and $\beta $ and core hole states $m$ and $n$ is assumed.}
\label{sigmae}
\end{figure}

\begin{figure}[b]
\psfig{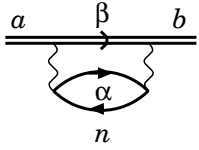}
\caption{Second-order diagram for the positron self-energy
($\hat \Sigma_p$ operator). Double line denotes positron states.}
\label{sigmap}
\end{figure}

The two-particle electron-positron wave function is given by the CI expansion,

\begin{equation}
\Psi({\bf r}_e,{\bf r}_p)=\sum_{i,j}C_{ij} \psi_i({\bf r}_e) \phi_j({\bf r}_p),
\label{c_i}
\end{equation}
where $\psi_i$ and $\phi_j$ are the electron and positron orbitals
respectively. The expansion coefficients $C_{ij}$ are to be determined by the
diagonalization of the matrix of the effective CI Hamiltonian acting in the
Hilbert space of the valence electron and the positron,
\begin{eqnarray}\label{HCI}
H_{\rm eff}^{\rm CI} &=& \hat h_e+\hat h_p + \hat h_{ep}, \nonumber \\
  \hat h_e &=&  c\bbox{\alpha p} + (\beta-1)mc^2
     - \frac{Ze^2}{r_e} + V_d^{N-1} - \hat V_{exch}^{N-1} + 
	\hat \Sigma_e, \nonumber \\
  \hat h_p  &=&  c\bbox{\alpha p} + (\beta-1)mc^2
    + \frac{Ze^2}{r_p} - V_d^{N-1} + \hat \Sigma_p, \\
\hat h_{ep} &=& - \frac{e^2}{|{\bf r}_e -{\bf r}_p|} - \hat \Sigma_{ep},
\nonumber 
\end{eqnarray}
where $\hat h_e$ and $\hat h_p$ the effective single-particle Hamiltonians of
the electron and positron, and $\hat h_{ep}$ is the effective electron-positron
two-body interaction. Apart from the relativistic Dirac operator, $\hat h_e$
and $\hat h_p$ include the direct and exchange Hartree-Fock potentials of the
core electrons, $V_d^{N-1}$ and $V_{exch}^{N-1}$, respectively. The additional
$\hat \Sigma$ operators account for correlations involving core electrons
(see \cite{Kozlov} for a detailed discussion). We calculate $\hat \Sigma $
using the second-order MBPT in the residual Coulomb interaction.
$\hat \Sigma_e$ describes the interaction between the valence electron and the
electrons of the core. All four second-order diagrams for the $\hat \Sigma_e$
are presented in Fig.~\ref{sigmae}. $\hat \Sigma_p$ is the 
correlation interaction between the positron and the core. In the second-order
$\hat \Sigma_p$ is represented by a sole digram in Fig.~\ref{sigmap}.
Both operators are often called correlation {\em potentials}, because
these {\em non-local} operators can be included into the equations for the 
single-particle orbitals together to the Hartree-Fock potential.
$\hat \Sigma_e$ and $\hat \Sigma_p$ are energy-dependent operators, which are
different for the electron and the positron. They are calculated separately
for each partial wave, ($s_{1/2}$, $p_{1/2}$, $p_{3/2}$, etc.). However, at
large distances both operators have the same asymptotic behaviour,
\begin{eqnarray}\label{alpha}
\Sigma _e({\bf r},{\bf r}'),\quad \Sigma _p({\bf r},{\bf r}') \simeq
	- \frac{\alpha e^2}{2r^4} \delta({\bf r} - {\bf r}'),
\end{eqnarray}
where $\alpha $ is the dipole polarizability of the atomic core. This
asymptotic form comes from the dipole contribution of the first diagram in
Fig.\ref{sigmae} for the electron, and diagram in Fig.\ref{sigmap} for the
positron. Formula (\ref{alpha}) with some empirical cut-off at small distances
is often used as an approximation for the correlation potentials, and is
usually called `polarization potential'.

% It is clear however that this approximation can be only very crude.

\begin{figure}[b]
\psfig{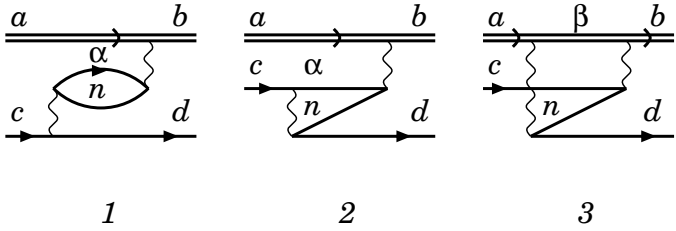}
\caption{Screening of the positron-electron Coulomb interaction
($\hat \Sigma _{ep}$ operator).}
\label{sigmaep}
\end{figure}

$\hat \Sigma_{ep}$ is another type of correlations between the external
particles and and core electrons. It can be described as screening of Coulomb
interaction between the external electron and positron by the core electrons.
There are in all six second-order diagrams for $\hat \Sigma_{ep}$. Three of
them are shown in Fig.~\ref{sigmaep}. The other three can be obtained from
them by mirror reflection with respect to the vertical axis.
% Note that we have no ``subtraction diagrams'' for all three
% $\Sigma$-operators. This is because we have the same atomic core in both
% Hartree-Fock and CI calculations (see \cite{Kozlov} for the discussion on
% subtraction diagrams).
When the electron and the positron are well outside the atomic core
$\hat \Sigma_{ep}$ is given by the following asymptotic expression,
\begin{eqnarray}\label{alphas}
 \Sigma_{ep} ({\bf r}_e,{\bf r_p}) \simeq \frac{\alpha e^2 {\bf r}_e \cdot
{\bf r}_p} {r_e^3 r_p^3} .
\end{eqnarray}
Similarly to Eq. (\ref{alpha}), this formula is often used to construct
rough approximations for $\hat \Sigma_{ep}$. Such potentials are
called `di-electronic correction', or `two-body polarization potential'.

Diagrammatic expansions in Figs. \ref{sigmae}, \ref{sigmap} and \ref{sigmaep}
enable one to include valence-core correlations in an {\em ab initio} manner.
To increase the accuracy of the calculations higher-order contributions to
$\hat \Sigma $ can be taken into account effectively, by introducing a
numerical factor before $\hat \Sigma $. For example, the coefficient for
$\hat \Sigma _e$ can be chosen by fitting the energies of the neutral atom 
states to the experimental data. In doing so the important non-local
structure of the operators is preserved.

\subsection{Basis set}

We use $B$-spline basis functions \cite{deBoor} to calculate the diagrams
for $\hat \Sigma$ and to construct the single-particle orbitals for the
CI expansion (\ref{c_i}). For this purpose the atomic system is confined to
a cavity of radius $R$, and the wave functions are set to zero at $r= R$.
For a sufficiently large $R$ the error introduced by this boundary condition
is very small for atomic-size binding energies, $\sim \exp (-2\kappa R)$,
where $\kappa $ is related to the binding energy as $\epsilon _B=
\kappa ^2\hbar ^2/2m$. However, for weakly bound states, e.g. those of
the positron with the atom, this error has to be considered more carefully
(see below). The interval $[0,R]$ is divided into a number of segments and
$B$-splines are constructed on them as piecewise polynomials of a certain
degree. They are bell-shaped overlapping smooth functions. With an appropriate
choice of the radial mesh they can approximate atomic wave functions to a very
high precision. Note that it is not convenient to use $B$-splines directly in
CI or MBPT calculations because of their non-orthogonality. Instead, we use
their  linear combinations which are eigenstates of the single-particle
Hartree-Fock Hamiltonian. This ensures orthogonality, allows to separate core
and valence states and improves convergence, since only a relatively small
number of lower Hartree-Fock eigenstates are sufficient for the convergence
of the CI calculation. This also means that while we use the same $B$-splines
for the electron and positron states the resulting single-particle basis
states are different, because the Hartree-Fock Hamiltonians for the electrons
and positrons are different. Another advantage of the use of $B$-splines is
that the convergence can be controlled by the cavity radius $R$ (its reduction
leads to a more rapid convergence), while its effect on the energy is taken
into account analytically.

\subsection{Effect of finite cavity size}\label{se:R}

The choice of the cavity radius $R$ (see above) is dictated by a compromise
between the convergence rate and the required accuracy of the calculations.
On one hand, the radius must be large enough to accommodate the wave function
of the state under investigation, e.g., the positron-atom bound state. On the
other hand, smaller radii mean faster convergence, both with respect to the
number of radial orbitals and, which is especially important for positron-atom
calculations, to the number of angular harmonics. This effect is very strong
since convergence is determined by the cavity volume which is proportional to
$R^3$, and having a smaller radius means that one needs fewer basis states
to describe the wave function.

The problem of convergence is crucial for the positron-atom interaction. As
discussed in the Introduction, the positron tends to form virtual Ps with the
external atomic electron \cite{Dzuba95,Dzuba:93,Gribakin:94}. The positronium
radius $r_{\rm Ps}\sim 2a_0$ can be small compared to the characteristic size
of the positron-atom bound state wave function, $r\sim 1/\kappa \gg a_0$, where
$a_0$ is the Bohr radius. To describe Ps at large separations from the atom
expansion (\ref{c_i}) needs to be extended to very high values of angular
momentum $L$ and principal quantum number $n$ to account accurately for the
virtual Ps formation. This problem is well known in positron-atom scattering
calculations, see e.g. \cite{Bray:93}. Smaller cavity radii force virtual Ps
to be at smaller distances, thereby improving the convergence significantly.
However the energy of the system is affected. Therefore, the convergence and
the accuracy of the calculation can be really improved only if the effect of
a finite-radius cavity on the energy is taken into account.

To consider the effect of cavity on the energy of the system let us consider
the problem of a particle weakly bound in an $s$ state by a finite-range
potential. `Weakly bound' here means that the binding energy is much smaller
than the typical scale of the potential. This is definitely true for
positron-atom bound states whose binding energy is much smaller than 1 eV.
To determine the radial wave function $\chi (r)$ at large distances it is
sufficient to impose on it a boundary condition
\begin{equation}\label{ata}
\left. \frac{1}{\chi } \frac{d\chi }{dr}\right| _{r=a} =-\kappa ~,
\end{equation}
at the outer radius $r=a$ of the potential well \cite{Landau:77}. The $\kappa $
parameter is related to the energy of the bound state $\varepsilon
=-\kappa ^2\hbar ^2/2m $, and determines the asymptotic form of the 
wave function, $\chi (r)\simeq A e^{-\kappa r}$.

The boundary condition
is unchanged when we place the system in the cavity of finite radius $R$,
$R>a$, provided the energy of the bound state is still small. However, the
wave function must now turn into zero at the cavity radius, $\chi (R)=0$.
This shifts the energy of the weakly bound state up from $\varepsilon $
to some other value $\varepsilon _R$, which depends on the radius of
the cavity. The Schr\"odinger equation for $a<r<R$, where the potential is
vanishingly small, is
\begin{equation}\label{Sch}
	\frac{\hbar ^2}{2m}\frac{d\chi ^2}{dr^2}+\varepsilon _R \chi (r)=0,
\end{equation}
After solving it with boundary conditions (\ref{ata}) and $\chi(R) = 0$, one
obtains a negative eigenvalue, $\varepsilon _R =-\kappa _R^2\hbar ^2/2m$,
where
\begin{equation}\label{elt0}
	\kappa = \kappa _R/ \tanh [\kappa _R(R-a)],
\end{equation}
if $R$ is not too small, $R-a>\kappa ^{-1}$.
As one can see, for $R\rightarrow \infty $ the solution of Eq. (\ref{elt0}),
$\kappa _R$, approaches its asymptotic value $\kappa $, and the energy
in the cavity $\varepsilon _R\rightarrow \varepsilon $. For a smaller cavity
radius the eigenvalue becomes positive, $\varepsilon _R =k_R ^2\hbar^2/2m$,
where $k_R$ is found from
\begin{equation}\label{egt0}
	\kappa = k_R / \tan [k_R (R-a)].
\end{equation}
This means that the state which is bound may appear as unbound due to
the effect of the cavity. Equation (\ref{egt0}) is valid for $k_R (R-a)
<\frac{\pi }{2}$. Otherwise, $\kappa <0$, and the energy is too high, 
so that it remains positive even when the cavity wall is removed.

Equations (\ref{elt0}) and (\ref{egt0}) can be used to find the
infinite-cavity energy $\varepsilon =-\kappa ^2\hbar ^2/2m$ from the energy
$\varepsilon _R$ calculated for the finite cavity radius $R$. It is important
that these formulae are insensitive to the detailed shape of the atomic
potential, and depend only on the atomic radius $a$. The value of $a$ can
be estimated from the position of the classical turning point $r_c$, in the
potential for an external atomic electron,
\begin{displaymath}
\frac{e^2}{r_c}=I=\frac{e^2}{2a_0\nu ^2},
\end{displaymath}
where $\nu $ is the effective quantum number of the valence electron. Beyond
the turning point the valence electron's wave function decreases exponentially,
as $\exp (-r/a_0\nu )$. Therefore, a resonable estimate for is
\begin{equation}\label{a}
a=r_c+a_0\nu =(2\nu ^2+\nu )a_0
\end{equation}
For copper ($I=0.28349$ a.u., $\nu =1.33$) this gives $a\approx 5a_0$. A
more accurate value of $a$ can be found by applying Eq. (\ref{elt0}) to
two bound-state calculations performed with two different cavity radii $R$.
The uncertainly in the value of $a$ is in fact unimportant, as long as we
consider weakly bound states for which $\kappa a\ll 1$.

Note that the wave function is also affected by the finite cavity size. 
This should be taken into account in calculations of the annihilation rate
and other matrix elements. The annihilation rate is proportional to 
the probability of finding the positron close to the atom. For $ a \ll 
R$ the wave function at $r \lesssim a$ is affected via normalization only.
The change of the normalization can be found by comparing the normalization
integral for $r >a$ calculated numerically within the cavity,
\begin{displaymath}
	\int _a ^R \chi^2(r)\,dr ~,
\end{displaymath}
with the analytical value
\begin{displaymath}
\int_a^\infty e^{-2\kappa  r}\,dr =
	\frac{1}{2\kappa }e^{-2\kappa a}.
\end{displaymath}

%=======================================================================

\section{Results and discussion} \label{results}

\subsection{Copper negative ion}\label{se:Cu-}

To test the method and find out what accuracy can be achieved
we first apply it to the copper negative ion. This is an effective
two-particle problem technically very similar to the positron-copper
interaction considered above. It should be mentioned that for Cu$^-$
only the electron $\hat \Sigma $-operator is involved (Fig. \ref{sigmae}),
and for the screening of the electron-electron interaction, instead of
the diagrams on Fig. \ref{sigmaep}, one must use similar diagrams presented
in \cite{Kozlov} (Fig.4). The results of calculations for Cu and Cu$^-$
are presented in table \ref{Cum} together with the experimental
values. The energies are given with respect to the Cu$^+$ core.
The accuracy of the Hartree-Fock approximation is very poor. The 
binding energy of the $4s$ electron in neutral Cu is underestimated by
about 20\%, while the negative ion Cu$^-$ appears altogether unbound
(its energy lies above that of the neutral atom).
The inclusion of core-valence correlations ($\hat \Sigma $) does improve
the energy of the neutral atom, but the negative ion is still not bound.
The standard CI method, in contrast, takes into account the valence-valence
correlations, while neglecting the core-valence correlations. It does
produce binding for the negative ion, but the binding energy is almost two
times smaller than the experiment value.
Only when both core-valence and valence-valence correlations are
included the accuracy improves significantly. It is equal to 2.6\%
for the ionization potential of the neutral atom and 10\% for the electron
affinity, which is quite good for a relatively simple {\em ab initio} 
calculation. The remaining discrepancy is mostly due to third and higher-order
correlation corrections in $\hat \Sigma $, since the configuration expansion
for Cu$^-$ converges rapidly, and the corresponding error is small.

To simulate the effects of higher-order terms in $\hat \Sigma $ and thus
further improve the accuracy of calculations we introduce numerical factors
before the $\hat \Sigma_e$ operators to fit the lowest $s$, $p$ and 
$d$ energy levels of the neutral copper atom. These factors are $f_s = 1.17$,
$f_p = 1.42$ and $f_d = 1.8$ in the $s$, $p$ and $d$ channels, 
respectively.
Table \ref{Cum} shows that these factors also significantly improve the
calculated electron affinity. It is natural to assume that the same
procedure should work equally well for the positron-atom problem.

Results of other calculations of the electron affinity of copper are
presented in table \ref{others}. Note that only a coupled-cluster (CC) method
produces a result more accurate than ours.
% This is another indication
%that the CC method is adequate for the positron-atom problem.
It is interesting to mention among other results the results by Mitroy
and Ryzhikh \cite{Mitroy98b,Mitroy99b} who calculated Cu$^-$ for 
the same purpose as we do, i.e., to gauge the accuracy of their method for
the positron-atom problem.
Their first result for electron affinity to copper, 0.921 eV, was obtained 
by the stochastic variational method, while another result 0.916 eV was
achieved in the standard CI calculation. Both methods are variational
in nature and differ basically by the form of the trial two-electron 
wave function. Since the two results agree well with each other, good
convergence has probably been achieved in both methods. However there is
a noticeable discrepancy between their result and the experimental
electron affinity value. From our point of view the
most probable source of this discrepancy is approximate treatment
of the $\hat \Sigma$ operator of the valence-core interaction.
In their works Mitroy and Ryzhikh use approximate expressions for the
core polarization potentials, based on asymptotic formulae (\ref{alpha})
and (\ref{alphas}), which include only dipole core polarization in
the local energy-independent form. Note again that the actual $\hat \Sigma$
operator is energy dependent. It is different for different angular
momenta, and for the electron and positron, while approximate expression
(\ref{alpha}) is always the same. Note also that the screening operator
$\hat \Sigma_{ep}$ depends not only on the states involved but also on the
multipolarity of the Coulomb integral. Approximate formula (\ref{alphas})
describes the dipole part of screening only, however, other 
Coulomb multipoles are also screened. Even though the largest 
contribution to screening comes from the dipole term, monopole
and quadrupole screening can not be neglected. For example, 
monopole screening directly contributes to the diagonal
Hamiltonian matrix elements in important configurations like $4s^2$ in
Cu$^-$, while dipole screening affects only the off-diagonal matrix
elements.

\subsection{Positron binding to copper}

The binding energy of Cu$^-$ is about 0.045 a.u. It corresponds
to a bound-state parameter $\kappa \approx 0.3$, and the cavity does not
have a  noticeable effect on the calculated energies of Cu$^-$, let alone Cu.
The relative error introduced by the cavity can be easily estimated from 
Eq. (\ref{elt0}), and even for a moderate $R=15a_0$ it does not exceed 0.1\%
for the electron affinity.

\begin{figure}[b]
\psfig{file=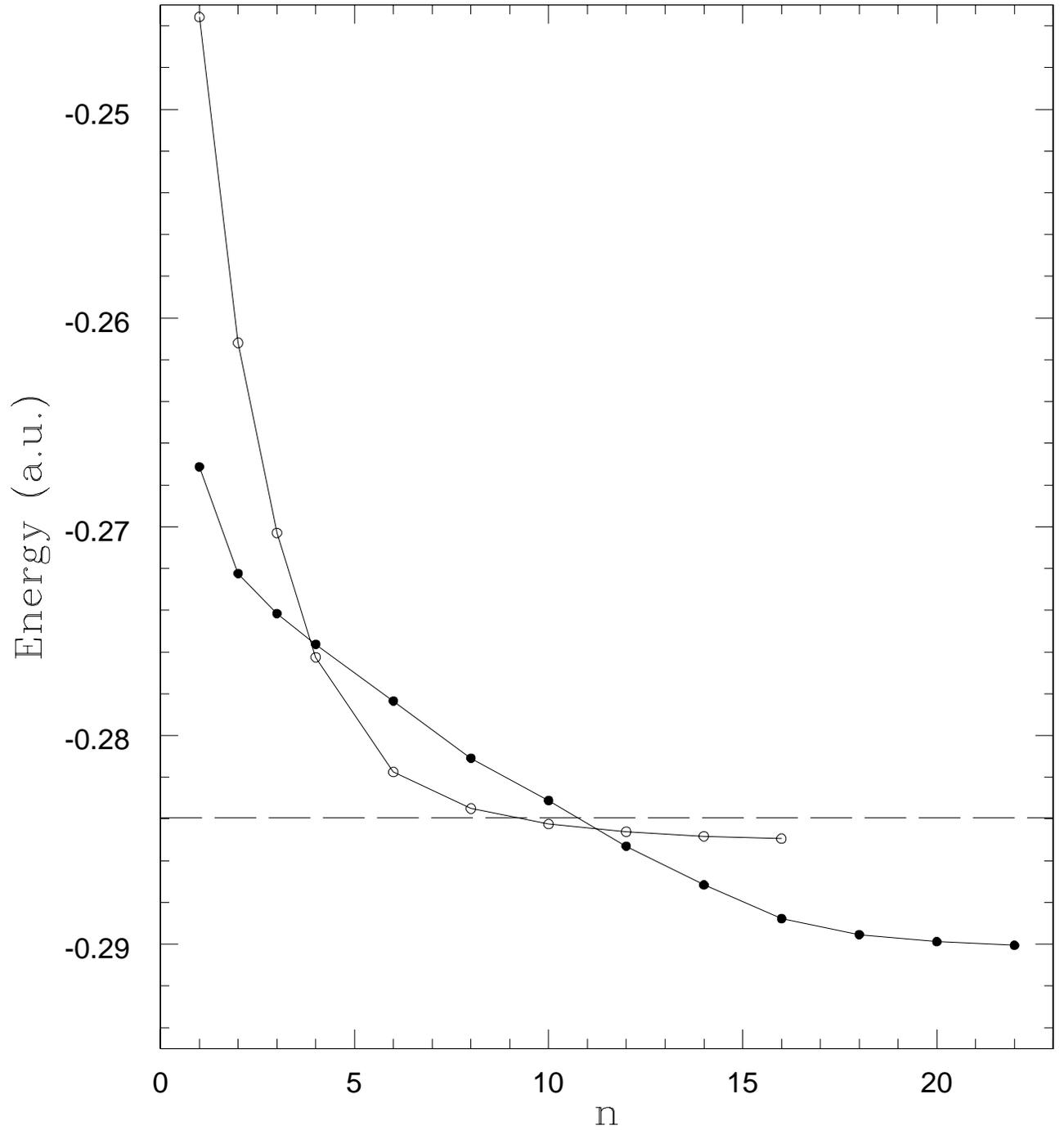, clip=}
\caption{Energy of Cu$e^+$ as a function of the number of radial electron
and positron basis functions in each partial
wave ($L_{\rm max}= 10$). Open circles are for $R=15a_0$, and solid ones
for $R=30a_0$.}
\label{f5}
\end{figure}

\begin{figure}[b]
\psfig{file=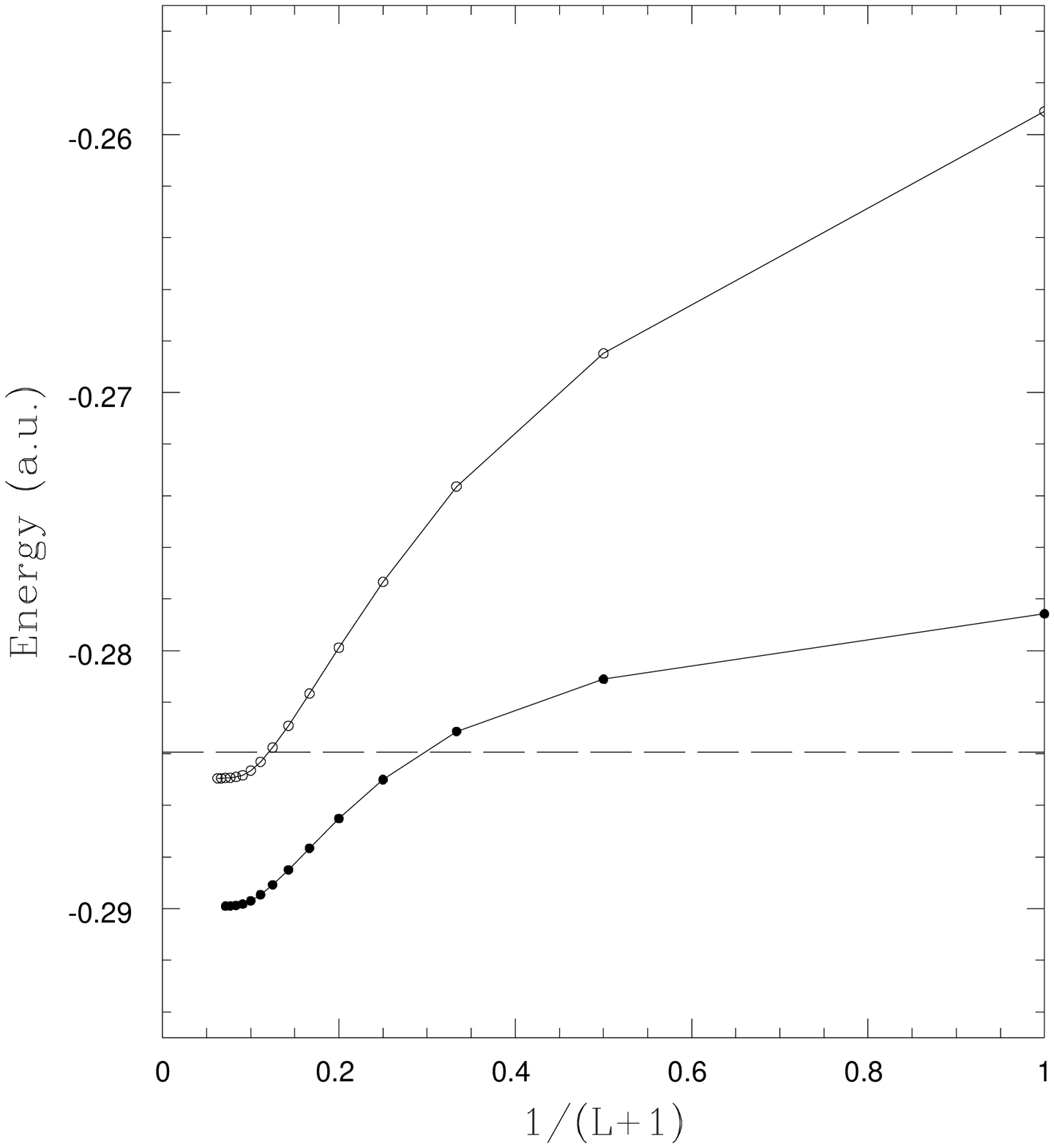, clip=}
\caption{Energy of Cu$e^+$ as a function
of maximal orbital momentum of the electron and positron orbitals in the CI
expansion. Open circles are for $R=15a_0$, and solid ones for $R=30a_0$.}
\label{f6}
\end{figure}

For the positron bound state the situation is different. As indicated
by the calculation of Ref. \cite{Mitroy98b}, the $\kappa $ value for
the Cu-$e^+$ bound state is about 0.1. This is why we have performed
the calculation of the positron-atom bound state using two different cavity
radii, $R=15a_0$ and $R=30a_0$, to make sure that convergence is really
achieved. The convergence pattern
with respect to the number of basis states used is illustrated in Figs.
\ref{f5} and \ref{f6}. Both plots show the energy of the
electron-positron pair moving in the field of Cu$^+$, with respect to the
energy of the Cu$^+$ ion (in atomic units). Empty circles correspond to
$R=15a_0$, while solid ones correspond to $R=30a_0$.
Dashed line shows the ground-state energy of the neutral copper atom.
The positron-atom state is bound when its energy is below the dashed line.
Fig. \ref{f5} shows the electron-positron energy of Cu-$e^+$ 
as a function of the number of radial basis functions in each electron
and positron partial wave, $n$. The total number of partial waves is fixed 
by $L_{\rm max} =10$. Note that convergence is visibly faster for the
smaller cavity radius. For $R=15a_0$ saturation begins at $n \approx = 10$
while for  $R=30a_0$ the same level of saturation can be seen only at
$n \approx = 18$.  Fig. \ref{f6} shows the Cu-$e^+$ energy
as a function of the number of partial waves included, while the number of
radial wave functions in each wave is fixed at $n=16$ for  $R=15a_0$
and $n=22$ for $R=30a_0$. Saturation can be clearly achieved for both
radii at $L_{\rm max} \gtrsim 10$. The difference in energy at the last
(lowest) points for $R=15a_0$ and $R=30a_0$ in both figures is the 
effect of a finite cavity radius. It shifts the energy obtained in the
$R=15a_0$ calculation up with respect to the $R=30a_0$ result.
This effect can be easily taken into account using
the formulae presented in section \ref{se:R}. It turns out that the results
for both cavity radii coincide, i.e., yield identical $\kappa $ from 
Eq. (\ref{elt0}), for the atomic potential radius of $a=5.5a_0$. The final
binding energy obtained is 0.0062 a.u., or 170 meV. This should be compared to
the result of Ryzhikh and Mitroy \cite{Mitroy98b}, which is 0.005518 a.u.
or 150 meV. From the discussion of the accuracy of calculations which
follows we conclude that the difference between two results is
within the accuracy of both methods. A similar value is achieved in
the CI calculation \cite{Mitroy99b}, which used 14 radial orbitals
in each partial wave up to $L_{\rm max}=10$, after extrapolation to
$L_{\rm max}=\infty $. However, the latter procedure has considerable
uncertainties.

There are several factors which affect the accuracy of our calculations.
\begin{itemize}
\item The accuracy of calculation of $\hat \Sigma$ and contributions of
higher-order correlations. This can be estimated by comparing calculations 
with and without the fitting parameters, as discussed in section \ref{se:Cu-}.
The introduction of the fitting parameters for the electron part of
the correlation
operator $\hat \Sigma_e$ reduces the binding energy by about 0.0009 a.u. 
However, the relevant uncertainty must be considerably smaller.
Firstly, we saw that the use of fitting parameters really improves the
calculated electron affinity of copper. We should expect the same effect for
the positron binding energy. Secondly, the effects of the fitting parameters
on the electron and positron operators $\hat \Sigma_e$ and $\hat \Sigma_p$
largely cancel each other.
\item Incompleteness of the basis set. We have seen from Figs \ref{f5} and
\ref{f6} that the level of convergence  achieved  is very high and 
the corresponding uncertainty is small. Nevertheless, there is a hidden
uncertainty related to the radial coordinate mesh used, the number of splines
and other parameters which determine the details of the numerical procedure.
Varying these parameters shows that their effect on the binding energy 
does not exceed 0.001 a.u., when estimated conservatively.
\item Finite cavity radius. This effect on the binding energy calculated
at $R=30a_0$ is very small ($\sim$ 0.0001 a.u.). Since the results for
$R=15a_0$ and $R=30a_0$ coincide for very reasonable value of the 
positron-atom potential radius $a=5.5a_0$, it is reasonable to believe that
the corresponding uncertainty is very small too.
\end{itemize}

Note that the difference between our calculated electron affinity of
copper and the experimental value is 0.00066 a.u. If this value is 
compared with the numbers presented above, it is evident that it also
gives a reasonable estimate of the accuracy of the calculation
of the positron-copper binding energy (about 10\%). Since the accuracy of
calculations by Mitroy and Ryzhikh \cite{Mitroy98b} is not discussed in their
paper the only thing we can do to estimate it is to use the same approach.
Their best result for Cu$^-$ differs from the experimental value by
0.0116 a.u. (20\% of the positron binding energy). If we adopt this value as
the uncertainty of their result
for the positron binding by copper, we see that the two results for 
Cu-$e^+$ bound state agree with each other within the accuracy of the
methods.

%*************************************************************************  

%---------------------------------------------
%**********************************************************************   
\begin{table}
\caption{Ground state energies of Cu and Cu$^-$ calculated using different
approximations (a.u.).}
\label{Cum}
\begin{tabular}{lddd}
 & Cu & Cu$^-$ & Electron affinity \\
\hline
RHF\tablenotemark[1]                  & $-$0.23830 & $-$0.20309 & $-$0.03521 \\
RHF + $\Sigma$\tablenotemark[2]       & $-$0.27672 & $-$0.27280 & $-$0.00392 \\
CI\tablenotemark[3]                     & $-$0.23830 & $-$0.26424 &  0.02594 \\
CI +$\Sigma$\tablenotemark[4]           & $-$0.27672 & $-$0.31802 &  0.04130 \\
CI + $f \times \Sigma$\tablenotemark[5] & $-$0.28394 & $-$0.32869 &  0.04475 \\
Experiment\tablenotemark[6]             & $-$0.28394 & $-$0.32935 &  0.04541 \\

\end{tabular}
\tablenotetext[1]{Relativistic Hartree-Fock; a single-configuration
approximation, no core-valence correlations are included.}
\tablenotetext[2]{Single-configuration approximation, core-valence 
correlations are included by means of MBPT.}
\tablenotetext[3]{Standard CI method.}
\tablenotetext[4]{CI+MBPT method, both core-valence and valence-valence
correlations are included.}
\tablenotetext[5]{$\Sigma$ for $s$-wave is taken with factor $f=1.18$ to
fit the Cu ground state energy.}
\tablenotetext[6]{References \cite{Moore,Bilodeau}.}
\end{table}

\begin{table}
\caption{Electron affinities of Cu (eV). Comparison with other calculations
and experiment.}
\label{others}
\begin{tabular}{dll}
Affinity & Ref. & Method\\
\hline
\multicolumn{3}{c}{Theory} \\
1.06 & \cite{Walch} & Nonrelativistic MR CI calculations \\
1.01 & \cite{Marian} & MR CI calculations in the DK no-pair formalism \\
1.199 & \cite{Bowmaker} & \\
1.236 & \cite{Neogrady} & Relativistic coupled cluster method \\
0.921 & \cite{Mitroy98b} & Nonrelativistic stochastic variational method\\
0.916 & \cite{Mitroy99b} & Nonrelativistic CI method\\
1.218 & \multicolumn{2}{c}{Present work} \\
\multicolumn{3}{c}{Experiment}\\
1.226 & \cite{Hotop}& \\
1.2358& \cite{Bilodeau}& \\
\end{tabular}
\end{table}
%####################################################################
\widetext
\newpage
\input psfig
\psfull

%------------------------------------------------------------------
%------------------------------------------------------------------
%------------------------------------------------------------------
%\begin{figure}[b]
%\psfig{file=pot.eps, clip=}
%\caption{Model potential to study the effect of the finite box size.}
%\label{pot}
%\end{figure}
%------------------------------------------------------------------
%------------------------------------------------------------------
%------------------------------------------------------------------
%\newpage
%------------------------------------------------------------------
%####################################################################

\begin{thebibliography}{25}

%%%%%%%%%%%%%%%%%e+ & Atom bound states %%%%%%%%%%%%%%%%%%%%%
\bibitem{aronson} I. Aronson, C. J. Kleinman and L Spruch, Phys. 
Rev. A {\bf 4}, 841 (1971).

\bibitem{gertler} F. H. Gertler, H. B. Snodgrass, and L. Spruch,
Phys. Rev. {\bf 172}, 110 (1968).

\bibitem{clary}D. C. Clary, J. Phys. B {\bf 9}, 3115 (1976).

\bibitem{ward}S. J. Ward, M. Horbatsch, R. P. McEachran, and A. D.
Stauffer, J. Phys. B {\bf 22}, 3763 (1989).

\bibitem{szmyt}R. Szmytkowski, J. Phys. II France {\bf 3}, 183 (1993);
R. Szmytkowski, Acta Physica Polonica A {\bf 84}, 1035 (1993).

\bibitem{Dzuba95}
V. A. Dzuba, V. V. Flambaum, G. F. Gribakin, and W. A. King, Phys. Rev. A
{\bf52}, 4541 (1995).

\bibitem{Mitroy97}  %Lie+
G. G. Ryzhikh and J. Mitroy, Phys. Rev. Lett. {\bf 79}, 4124 (4124).

\bibitem{Strasburger98}
K. Strasburger and H. Chojnacki, J. Chem. Phys. {\bf 108}, 3218
(1998).

\bibitem{small} % Na, He e+
G. G. Ryzhikh and J. Mitroy, J. Phys. B. {\bf 31}, L265 (1998);
{\bf 31} 3465 (1998);

\bibitem{Yuan98} % Li Nae+ : adiabatic hyperspherical calculation
J. Yuan, B. D. Esry, T. Morishita, and C. D. Lin, Phys. Rev. A {\bf 58},
R4 (1998).

\bibitem{Mitroy98} %Mge+
G. G. Ryzhikh and J. Mitroy, J. Phys. B. {\bf 31}, L401 (1998).

\bibitem{RyMitVar98}
G. G. Ryzhikh, J. Mitroy and K. Varga, J. Phys. B. {\bf 31}, 3965 (1998).

\bibitem{Mitroy98b}   %Cue+
G. G. Ryzhikh and J. Mitroy, J. Phys. B. {\bf 31}, 4459 (1998).

\bibitem{Mitroy98c}   %Age+
G. G. Ryzhikh and J. Mitroy, J. Phys. B. {\bf 31}, 5013 (1998).

\bibitem{Mitroy99a} % Zne+
J. Mitroy and G. G. Ryzhikh, J. Phys. B. {\bf 32}, 1375 (1999).

\bibitem{Mitroy99b}   %Cue+ CI
J. Mitroy and G. G. Ryzhikh, J. Phys. B. {\bf 32}, 2831 (1999).

\bibitem{Dzuba:93}
V. A. Dzuba, V. V. Flambaum, W. A. King, B. N. Miller, and O. P. Sushkov,
Phys. Scripta T {\bf 46}, 248 (1993).

\bibitem{Gribakin:94}
G. F. Gribakin and W. A. King, J. Phys. B {\bf 27}, 2639 (1994).

\bibitem{Dzuba:96}
V. A. Dzuba, V. V. Flambaum, G. F. Gribakin, and W. A. King, J. Phys. B
{\bf 29}, 3151 (1996).

%\bibitem{Lindgren} %CC
%	I.~Lindgren, J.~Morrison, {\it Atomic Many-Body Theory}
%	(Springer-Verlag, Berlin, Second edition, 1985).

\bibitem{Kozlov}
V. A. Dzuba, V. V. Flambaum, and M. G. Kozlov, Phys. Rev. A {\bf54}, 3948
(1996).

\bibitem{deBoor}
C. deBoor, {\it A Practical Guide to Splines} ( Springer, New York, 1978).

\bibitem{Sapirstein}
J. Sapirstein and W. R. Johnson, J. Phys. B {\bf29}, 5213 (1996).

\bibitem{Johnson}
V. A. Dzuba and W. R. Johnson, Phys. Rev. A {\bf 57}, 2459 (1998).

\bibitem{pol}A. A. Radtsig and B. M. Smirnov, {\em Parameters of
Atoms and Atomic Ions: Handbook} (Energoatomizdat, Moscow, 1986);
{\it CRC Handbook of Physics and Chemistry}, 69th edition, Editor-in Chief
R. C. Weast (Boca Raton, Florida, CRC Press, 1988).

\bibitem{Moore}
C. E. Moore, {\it Atomic Energy Levels}, Natl. Bur. Stand. Circ. No. 467
(U.S. GPO, Washington, DC, 1958), Vol. III.

\bibitem{Bray:93}
I. Bray and A. Stelbovics, Phys. Rev. A {\bf 48} 4787 (1993).

\bibitem{Landau:77}
L. D. Landau and E. M. Lifshitz, {\it Quantum Mechanics}, 3rd ed. (Pergamon
Press, Oxford, UK, 1977).

%%%%%%%%%%%%%%%%% Cu- %%%%%%%%%%%%%%%%%%%%%
\bibitem{Walch}
C. W. Bauschlicher Jr., S. P. Walch, and H. Partridge, Chem. Phys. Lett.
{\bf 103}, 291 (1984).
\bibitem{Marian}
C. M. Marian,  Chem. Phys. Lett. {\bf 173}, 175 (1990).
\bibitem{Bowmaker}
P. Schwerdtfeger and G. A. Bowmaker, J. Chem. Phys. {\bf 100}, 4487 (1994).
\bibitem{Neogrady}
P. Neogrady, V. Kello, M. Urban, and A. J. Sadrej, Int. J. Quantum Chem.
{\bf 63}, 557 (1997).
\bibitem{Hotop}
H. Hotop and W. C. Lineberger, J. Phys. Chem. Ref. Data {\bf 14}, 731 (1975).
\bibitem{Bilodeau}
R. C. Bilodeau, J. Phys. {\bf B}, 3885 (1998).
\end{thebibliography}
\end{document}